# Intertwined Topological and Magnetic Orders in Atomically Thin Chern Insulator MnBi$_2$Te$_4$

**Authors**: Dmitry Ovchinnikov[1*], Xiong Huang[2,3*], Zhong Lin[1*], Zaiyao Fei[1*], Jiaqi Cai[1], Tiancheng Song[1], Minhao He[1], Qianni Jiang[1], Chong Wang[4], Hao Li[5], Yayu Wang[6], Yang Wu[7], Di Xiao[4], Jiun-Haw Chu[1], Jiaqiang Yan[8,9§], Cui-Zu Chang[10§], Yong-Tao Cui[2§], Xiaodong Xu[1,11§]

**Affiliations:**
[1]Department of Physics, University of Washington, Seattle, Washington 98195, USA.
[2]Department of Physics and Astronomy, University of California, Riverside, California 92521, USA
[3]Department of Materials Science and Engineering, University of California, Riverside, California 92521, USA
[4]Department of Physics, Carnegie Mellon University, Pittsburgh, Pennsylvania 15213, USA
[5]School of Materials Science and engineering, Tsinghua University, Beijing, 100084, P. R. China
[6]Department of Physics, Tsinghua University, Beijing 100084, P. R. China
[7]Department of Mechanical Engineering, Tsinghua University, Beijing 100084, P. R. China
[8]Materials Science and Technology Division, Oak Ridge National Laboratory, Oak Ridge, Tennessee 37831, USA.
[9]Department of Materials Science and Engineering, University of Tennessee, Knoxville, Tennessee 37996, USA.
[10]Department of Physics, The Pennsylvania State University, University Park, Pennsylvania 16802, USA
[11]Department of Materials Science and Engineering, University of Washington, Seattle, Washington 98195, USA.
*These authors contributed equally to the work.
§Correspondence to yanj@ornl.gov; cxc955@psu.edu; yongtao.cui@ucr.edu; xuxd@uw.edu

**Abstract:** The interplay between band topology and magnetic order plays a key role in quantum states of matter. MnBi$_2$Te$_4$, a van der Waals magnet, has recently emerged as an exciting platform for exploring Chern insulator physics[1,2]. Its layered antiferromagnetic order was predicted to enable even-odd layer-number dependent topological states[3-5], supported by promising edge transport measurements[6-9]. Furthermore, it becomes a Chern insulator when all spins are aligned by an applied magnetic field. However, the evolution of the bulk electronic structure as the magnetic state is continuously tuned and its dependence on layer number remains unexplored. Here, employing multimodal probes, we establish one-to-one correspondence between bulk electronic structure, magnetic state, topological order, and layer thickness in atomically thin MnBi$_2$Te$_4$ devices. As the magnetic state is tuned through the canted magnetic phase, we observe a band crossing, i.e., the closing and reopening of the bulk bandgap, corresponding to the concurrent topological phase transition. Surprisingly, we find that the even- and odd-layer number devices exhibit a similar topological phase transition coupled to magnetic states, distinct from recent theoretical[3-5] and experimental[6-9] reports. Our findings shed new light on the interplay between band topology and magnetic order in this newly discovered topological magnet and validate the band crossing with concurrent measurements of topological invariant in a continuously tuned topological phase transition.

**Main text:**

Chern insulators, which are quantum Hall insulators without the formation of Landau levels, possess quantized Hall conductance with dissipation-free chiral edge states[10]. The interplay between topological physics and time reversal symmetry breaking is a prerequisite for the realization of the Chern insulator state[11]. So far, the Chern insulator state has been experimentally realized in magnetically doped topological insulator thin films/heterostructures[12-18] and the recently discovered twisted graphene moiré systems[19-21].

Recently, van der Waals magnet $MnBi_2Te_4$ has emerged as a new system to realize the Chern insulator state[1-9]. In contrast to existing systems[12-21], whose parent materials have no intrinsic magnetic order, $MnBi_2Te_4$ is an intrinsic layered antiferromagnet (AFM)[1,2]. The adjacent ferromagnetic (FM) monolayers with out-of-plane magnetization are AFM coupled, which can be further driven into a canted antiferromagnetic (cAFM) and FM states by applying a magnetic field (Fig. 1a). The coupling of different magnetic configurations to the topological states are predicted to lead to a variety of topological phenomena[3-5]. For instance, in the AFM ground state, an odd-layer sample possesses effectively one net layer of uncompensated magnetization and is predicted to host the QAH state, while the magnetization is fully compensated in an even-layer sample, which is expected to be an axion insulator[3-5]. Experimentally, using exfoliated atomically thin devices, gate tunable quantized Hall resistance with a Chern number $C = -1$ has been observed in the magnetic field-driven FM states[6-9] as well as at zero magnetic field in an odd layer number device[6].

These promising results point to atomically thin $MnBi_2Te_4$ as a desirable platform to study the correlation between band topology and magnetic order. So far, experiments have focused on edge transport, and the corresponding magnetic states are usually inferred from Hall measurements. Direct correlation between magnetic and topological orders has yet to been established. It is also unclear how the bulk electronic band structure evolves as the magnetic state is tuned, and how the even-odd layer number effects are manifested during the topological phase transition (TPT), which is unique to this layered AFM system.

In this work, we unravel the intertwined band topology and magnetic order as a function of $MnBi_2Te_4$ layer thickness. This is achieved by employing reflective magnetic circular dichroism (RMCD) to identify magnetic states, scanning microwave impedance microscopy (MIM) to image edge conduction and measure local bulk conductivity, and electrical transport to measure bulk resistance as well as the topological invariant via anomalous Hall traces. Our devices are fabricated using mechanically exfoliated $MnBi_2Te_4$ thin flakes (see Methods). Figure 1b shows atomic force microscopy micrograph of adjacent 6 and 7 septuple layer (SL) flakes. The linecut in Fig. 1c shows the thickness of the 6 SL flake. The single SL step between those two flakes is ~1.4 nm thick, consistent with prior reports [6-8].

Since the even- and odd-layer samples have compensated and uncompensated magnetic moments, respectively, the total magnetization should exhibit distinct magnetic-field dependence. We first investigate this even-odd effect by RMCD measurements (see Methods). Figure 1e illustrates the RMCD signal for flakes with thicknesses from 4 to 8 SL. Two major differences in RMCD signal between even and odd layers stand out. First, odd SL flakes have pronounced hysteresis loops and significant remanent RMCD signals in the AFM state, which results from the uncompensated magnetic moments, while even layers have vanishingly small RMCD signals and hysteresis loops, consistent with their zero net magnetization. Second, the critical magnetic field for the spin-flop transition between the AFM and cAFM states, labeled by $H_C$ with blue and red arrows in Fig. 1e, is about 1.8 T for even SL flakes, which is distinct from the value of ~3.8 T for odd SL flakes. Our RMCD results are qualitatively consistent with a recent preprint[22]. The larger $H_C$ for odd SL compared to the even ones can be explained by the Zeeman energy of the uncompensated magnetization in the former case, which stabilizes the AFM state. This even-odd layer dependent RMCD signal is well captured by our theoretical modeling of $M$ vs $B$ field (Fig. 1f, see Methods for calculation details) and appears to be a reliable way to distinguish between even and odd SL flakes.

We next investigate the correlation between magnetic states and the Chern number $C$ in both even and odd SL MnBi$_2$Te$_4$ devices. We show such a side-by-side comparison between a 6-SL (Figs. 2a-c) and a 5-SL (Figs. 2d-f) device. Overall, the RMCD signals of the devices are consistent with the freshly cleaved flakes (Fig. 1e). For the 6-SL device (6SL-1), RMCD signal in the AFM state nearly vanishes, although a tiny hysteresis loop exists (Fig. 2a), which does not appear in freshly cleaved 6-SL samples. We speculate that its formation is likely a result of uncompensated magnetization initiated by fabrication (Extended Data Fig. 2), which is often seen in 2D layered antiferromagnet CrI$_3$.[23] The odd layer device 5SL-1 has an expected hysteresis loop centered around $B = 0$ T (Fig. 2d). The sharp transition is the signature of a spin-flip transition between two magnetic states with the same net magnetic moment but opposite signs. The remnant RMCD signal is about 28% of that in the FM state, which is slightly larger than the expected value of 20%. The RMCD map under $B = 0$ T also shows a uniform signal (Fig. 2d inset), suggesting large magnetic domains[22,24].

The transport signal shows that the spin-flop fields, determined by the rising point of $R_{yx}$, are consistent with the RMCD measurements. For both devices, a Chern insulator state of $C = -1$ is observed in the magnetic field induced FM state when $B > 6$ T, made evident by quantized Hall resistance $R_{yx} = -h/e^2$ (Figs. 2b and 2e) and vanishing longitudinal resistance $R_{xx}$ (Figs. 2c and 2f). These are consistent with prior reports [6-9]. In the AFM state of the even layer device (6SL-1) at $B < \sim 2$ T, $R_{yx}$ is small while $R_{xx}$ is large. Although RMCD nearly vanishes, $R_{yx}$ has a small hysteresis loop with a coercive field that differs from that measured by RMCD. From our spatially resolved RMCD measurements, small hysteresis loops similar to those in $R_{yx}$ are observed for some spots near the sample edges, suggesting that the loops might be caused by imperfections near the edge of the device (Extended Data Fig. 2).

Unexpectedly, for the odd layer device (5SL-1) in the AFM state at $|B| < H_C \sim 3.8$ T, both $R_{yx}$ and its hysteresis loop nearly vanish, in stark contrast to the observed uncompensated magnetic moment and pronounced hysteresis in RMCD. In addition to the close-to-zero $R_{yx}$, $R_{xx}$ reaches a large value of about 7 $h/e^2$. Our results suggest that the AFM state in our 5-SL device is likely a trivial magnetic insulator in which a vanishing $R_{yx}$ is expected. We remark that although our transport results look similar to the prior reports, the assignment of even-odd is opposite[6-9]. This implies that in addition to atomic force microscopy and transport measurements, the identification of the exact magnetic states, such as by concurrent RMCD measurement on the same device, are needed for the interpretation of the even/odd layer number dependent topological effects.

After comparing the even-odd layer number effects in the FM and AFM states, we turn our attention to the evolution of bulk electronic structure during the magnetic field-driven TPT in the cAFM states. We find that both even- and odd-SL devices exhibit a similar band-crossing behavior, with subtle variations due to their differing magnetization vs $B$-field behaviors (see Extended Data Figs. 3 and 4 for data on two 5-SL devices). To probe the bulk states, we measure the local conductivity of the sample bulk by MIM. Using a 6-SL device (6SL-2) as an example, a metal tip is scanned across the sample and the complex impedance between the tip and sample is measured. The real and imaginary parts of the MIM response, MIM-Im and MIM-Re, are used to analyze the sample conductivity[25]. In particular, the MIM-Im channel, which characterizes the ability of the sample to screen the electric field from the tip, increases monotonically with the local conductivity. To avoid edge contribution, we measure the MIM signal of bulk by parking the tip near the center of the flake, ~ 4 μm away from the edge (Fig. 3a inset). Figure 3a shows the intensity plot of the MIM-Im signal, $\sigma_{MIM}$, as a function of $B$ and $V_{bg}$. In addition, we measure bulk resistance $R_{bulk}$ via transport using the contact configuration illustrated in the inset of Fig. 3b. We apply a small AC voltage bias between a pair of electrodes on opposite sides of the Hall bar while grounding all other electrodes. Since the edge conduction is grounded, the current mainly goes through the sample bulk (see Methods). The $V_{bg}$-$B$ map of bulk resistance $R_{bulk}$ is shown in Fig. 3b. To correlate the bulk states with the topological properties, the $V_{bg}$-$B$ maps of $R_{xx}$ and $R_{yx}$ are also measured (Figs. 3e and 3f).

These four maps of complementary quantities show striking similarities in the $V_{bg}$ and $B$ dependence, which reveals the evolution of the bulk electronic structure during the magnetic field induced TPT. Importantly, the local (Fig. 3a) and bulk (Fig. 3b) resistance maps match well, demonstrating the robustness of our measurements on the bulk electronic states. These two maps show that in both the AFM ($B < 2$ T) and FM ($B > 6$ T) regimes, there is a resistive state within a certain gate range (see also the line cuts of $R_{bulk}$ vs $V_{bg}$ at selected $B$ in Fig. 3c). These resistive states correspond to the range of $V_{bg}$ where the Fermi level is tuned inside the bulk bandgap (Fig. 3d). Upon comparison to the $R_{xx}$ and $R_{yx}$ maps, we can see that the bulk gap in the AFM state corresponds to the $C = 0$ insulating state, which has close-to-zero $R_{yx}$ and large $R_{xx}$, implying its trivial origin. In contrast, the highly insulating bulk state in the FM state at high fields coincides with the gate range for quantized $R_{yx}$ and vanishing $R_{xx}$ (Figs. 3e and 3f). Overlaying the contour of $R_{yx}$ at a threshold of 98% of $-h/e^2$ on all maps displays an excellent correlation among all four data sets. Furthermore, spatial MIM scans reveal the formation of conductive edge states in this regime (see the upper inset of Fig. 3a and Extended Data Fig. 5 for a complete MIM image set showing the evolution of spatial conductivity profiles near the edges). All these measurements mark the observation of the Chern insulator gap with $C = -1$ in the FM phase.

The cAFM phase connects the $C = 0$ insulating state in the AFM phase to the $C = -1$ Chern insulator in the FM phase. The $\sigma_{MIM}$ and $R_{bulk}$ maps show that as the magnetic field $B$ increases above ~2 T, the gate regime for the $C = 0$ insulating state shrinks while an incipient Chern insulator gap appears and increases, with a conductive state separating the two regimes. Specifically, as $B$ increases, one bulk conduction band of the AFM phase splits off and moves toward the bulk valence band. Eventually, it crosses the entire bandgap and merges with the bulk valence band at $B \sim 7$ T in the FM phase (Fig. 3d). Meanwhile, the Chern insulator gap increases and eventually fully opens, accompanying the gradual increase of $|R_{yx}|$ to the quantized value of $h/e^2$ and the decrease of $R_{xx}$ towards zero. Therefore, what we observe is the band crossing process, where canting of the magnetic moment is responsible for the energy shift of the crossing band via exchange interaction.

Finally, we investigate the temperature dependence of the band crossing and Chern insulator gap to examine their correlation with the formation of the magnetic order. Figure 4 illustrates the temperature dependence of the electronic properties in device 6SL-2. (Results of 5SL-1 with a similar temperature dependence are presented in Extended Data Fig. 6.) Our thin flakes of MnBi$_2$Te$_4$ typically have a magnetic ordering temperature $T_N \sim 23$ K (Extended Data Fig. 7).

Figure 4a shows the $R_{yx}$ and $R_{xx}$ as a function of $V_{bg}$ at $B = 9$ T at selected temperatures. We correlate the transport measurements with the band crossing feature probed by $\sigma_{MIM}$ and $R_{bulk}$ (Figs. 4b and 4c and see also Extended Data Fig. 8 for edge states measured at all temperatures). At $T = 20$ K and below, the band crossing occurs in the $B$ range between 2 T and 7 T. At $T = 30$ K, the splitting of the sub-band from the conduction band starts at a higher $B$ and the band shift becomes more gradual. At $B = 9$ T, the crossing branch touches the bulk valence band without a complete merge. There is a close-to-quantization region where $R_{yx}$ is still as high as 90% of $h/e^2$ and $R_{xx}$ remains low, indicating the persistence of the chiral edge state. At $T = 40$ K, the close-to-quantization region shrinks further with a maximum of $R_{yx} \sim 0.8\ h/e^2$, and the crossing band barely touches the bulk valence band at $B = 9$ T. The large Hall angle $R_{yx}/R_{xx}$ at temperatures above $T_N$ is similar to a recent report[8], which is consistent with the observed incipient but not fully developed Chern insulator gap. For $T = 50$ K and above, $R_{yx}$ significantly deviates from the quantized value and the crossing branch is completely detached from the bulk valence band at $B = 9$ T.

These results suggest that below $T_N$, the FM ordered moments induce a large exchange interaction in a conduction sub-band, which fully crosses the bulk valence band above 7 T, leading to a robust Chern insulator gap. While the long-range ordering of Mn moments disappears at temperatures slightly above $T_N$, sufficiently large magnetic fields can still align the moments to introduce an incipient Chern insulator state, albeit one much weaker than the state which exists below $T_N$. Our results highlight that to achieve a comprehensive understanding of the topological physics in this new Chern magnet, it is necessary to combine measurements of local and bulk electronic properties with a careful examination of magnetic states. With a multitude of possible topological phases and magnetic orders in MnBi$_2$Te$_4$ and related topological magnets[3-5,26-29], our work paves the way towards a full realization of their potential as a platform to study topological phase transitions and to engineer Chern numbers in Chern insulators via external control knobs.

**Methods:**

**Device fabrication:** MnBi$_2$Te$_4$ bulk crystals from Oak Ridge National Lab were grown out of a Bi-Te flux as previously reported. The commercial Mn pieces (Alfa Aesar, 99.95%) were cleaned to remove surface oxides by arc melting before use. Devices were fabricated inside an Ar-filled glovebox. We used standard scotch-tape exfoliation onto 285 nm thick SiO$_2$/Si substrates to obtain MnBi$_2$Te$_4$ flakes. Next, thin flakes were identified by a combination of optical contrast and atomic force microscopy imaging. We proceeded with standard electron beam lithography, employing 300 nm PMMA resist and thermal evaporation of Cr (5 nm) and Au (50 nm), followed by liftoff. The transport devices were covered by PMMA during all measurements to protect flakes from degradation. Note that PMMA covering may introduce spatial charge inhomogeneity, as revealed by our MIM measurements. To study layer dependent magnetic states by RMCD, the flakes presented in Fig. 1 were freshly exfoliated, sealed in a vacuum protected sample mount, and measured without PMMA covering. We used two crystal sources for this work. The device 5SL-2 was fabricated from crystals grown by Y. Wu. All other samples were fabricated from crystals grown by J. Yan.

**Transport measurements:** Transport measurements were performed in a PPMS (Quantum Design, 1.75 K/9 T). The magnetic field was applied perpendicular to the sample plane. Standard lock-in techniques with low excitation currents of 2~10 nA were used to measure $R_{xx}$ and $R_{yx}$ simultaneously. The $R_{xx}$ ($R_{yx}$) data shown in the main text were symmetrized (anti-symmetrized) as a function of the magnetic field to eliminate the effect of electrode misalignments. The raw $R_{xx}$ and $R_{yx}$ data of 6SL-1 sample are shown in Extended Data Fig. 1.

Bulk resistance measurements were performed by biasing a pair of electrodes on opposite sides of the flake with a small AC bias (typically between 100 and 250 µV) and measuring AC current with lock-in techniques, with other electrodes grounded. Notably, there are no direct edge channels connecting the biasing pair such that the majority of current flow occurs through bulk $MnBi_2Te_4$, rather than through the edge channels. An example of the configuration used in our measurements is shown in the inset of Fig. 3b. We note that ideally a Corbino geometry in which no edge channels are involved is needed to measure the true bulk resistance. Although edge channels are still indirectly connecting the contacts in our configuration, its validity to extract, at least semi-quantitatively, the bulk resistance can be justified by its correlation with the bulk MIM signal which only probes the conductivity locally.

**Microwave impedance microscopy measurements:** MIM measurements were performed in a home-built cryogenic scanning probe system. A small microwave excitation of ~0.1 µW at a fixed frequency in the range of 1 ~ 10 GHz was delivered to a chemically etched tungsten tip mounted on a quartz tuning fork. The reflected signal was analyzed to extract the demodulated output channels, MIM-Im and MIM-Re, which are proportional to the imaginary and real parts of the admittance between the tip and the sample, respectively. To enhance the MIM signal quality, the tip was excited to oscillate at a frequency of ~32 kHz with an amplitude of ~8 nm. The resulting oscillation amplitudes of MIM-Im and MIM-Re were then extracted using a lock-in amplifier to yield d(MIM-Im)/d$z$ and d(MIM-Re)/d$z$, respectively. The d(MIM)/d$z$ signals are free of fluctuating backgrounds and their behaviors are very similar to that of the standard MIM signals. In this paper we simply refer to d(MIM)/d$z$ as the MIM signal. For some samples used in MIM measurements, transport was monitored simultaneously whenever possible to correlate the MIM data with $R_{xx}$, $R_{yx}$ and $R_{bulk}$ measured by transport.

**Reflective magnetic circular dichroism measurements:** A material with non-zero magnetic moment in the out of plane direction may exhibit both (*i*) magnetic circular birefringence (MCB), which results in a phase difference between right-circularly polarized (RCP) light and left-circularly polarized (LCP) light, and (*ii*) magnetic circular dichroism (MCD), which induces an amplitude difference between RCP and LCP light. When linearly polarized light, which is an equal superposition of RCP and LCP light, is reflected from such material surface, the linear polarization is rotated by an angle known as the Kerr rotation from MCB and becomes elliptically polarized due to MCD. The former is employed to measure the magneto-optical Kerr effect (MOKE), while the latter results in reflective magnetic circular dichroism (RMCD). In our experiment, RMCD measurements were performed in a closed-cycle dry cryostat (attoDRY 2100) at the base $T$ = 1.6 K and magnetic fields up to 9 T. Linearly polarized light with a wavelength of 632.8 nm and a typical power of 0.5 – 1 µW was focused through an aspheric lens to form a ~2 µm beam spot on the sample surface with normal incidence. Experimental setup closely follows previous measurements of magnetic order in $CrI_3$ by means of MOKE and RMCD [23].

**Theoretical modelling of spin flop:** A model with interlayer exchange coupling, magnetic anisotropy, and external magnetic field is adopted to characterize the spin flop in $MnBi_2Te_4$ thin flakes. We model each SL layer as a macro-spin with the magnetization of *n*-th layer given by $\boldsymbol{M}_n = M_s(\sin\phi_n, 0, \cos\phi_n)$ and the total magnetic energy is written as $E = M_s[\sum_{n=1}^{N-1} J\cos(\phi_{n+1} - \phi_n) - \sum_{n=1}^{N} K\cos^2\phi_n - \sum_{n=1}^{N} B\cos\phi_n]$, where $N$ is the total number of SLs. The magnetic energy is minimized by a combination of global optimization (differential evolution) and local optimization (conjugate gradient method). The parameters $J$ and $K$ are fitted to the first experimental spin-flop fields and take values of $J = 1.76$ T and $K = 0.75$ T.

**Acknowledgements:** We thank David Cobden and Matthew Yankowitz for helpful discussions, Bevin Huang for initial assistance in RMCD measurements, Jonathan Kephart for advice on fabrication, Anna Isaeva and Zhiqiang Mao who separately provided crystals for test, and Jordan Fonseca and Eric Anderson for proofreading this manuscript. Research on topological properties of $MnBi_2Te_4$ is primarily supported as part of Programmable Quantum Materials, an Energy Frontier Research Center funded by the U.S. Department of Energy (DOE), Office of Science, Basic Energy Sciences (BES), under award DE-SC0019443. Magneto optical spectroscopy measurement is partially supported by DOE BES DE-SC0018171. Device fabrication and part of transport measurement are supported by AFOSR MURI program, grant no. FA9550-19-1-0390. The authors also acknowledge the use of the facilities and instrumentation supported by NSF MRSEC DMR-1719797. J.Y. acknowledges support from the U.S. Department of Energy, Office of Science, Basic Energy Sciences, Materials Sciences and Engineering Division. C.Z.C. acknowledges the support from ARO Young Investigator Program Award (W911NF1810198) and the Gordon and Betty Moore Foundation's EPiQS Initiative (Grant GBMF9063). X.H. and Y.T.C. acknowledge support from NSF under award DMR-2004701, the Hellman Fellowship award, and the seed fund from SHINES, an EFRC funded by DOE BES under award DE-SC0012670. H.L., Y. Wu, and Y. Wang were supported by NSFC 21975140, 51991313. X.X. and J.H.C. acknowledge the support from the State of Washington funded Clean Energy Institute. Computing time is provided by BRIDGES at the Pittsburgh supercomputer center (DMR200085) under the Extreme Science and Engineering Discovery Environment (XSEDE) supported by NSF (ACI-1548562).

**Author contributions:** X.X., Y.T.C., C.Z.C., J.Y., J.H.C., D.X. supervised the project. D.O., Z.L., Z.F., and J.C. fabricated devices, and performed transport and optical measurements, assisted by T.S., M.H., and Q.J.. X.H. performed MIM measurements. C.W. and D.X. provided theoretical support. J.Y. as well as H.L., Y.Wang, and Y.Wu provided and characterized bulk crystals separately. All authors contributed to data analysis. D.O., X.X., Y.T.C., C.Z.C., and D.X. wrote the paper with inputs from all authors.

**Competing Interests:** The authors declare no competing interests.

**Data Availability:** The datasets generated and/or analyzed during this study are available from the corresponding author upon reasonable request.

# Figures:

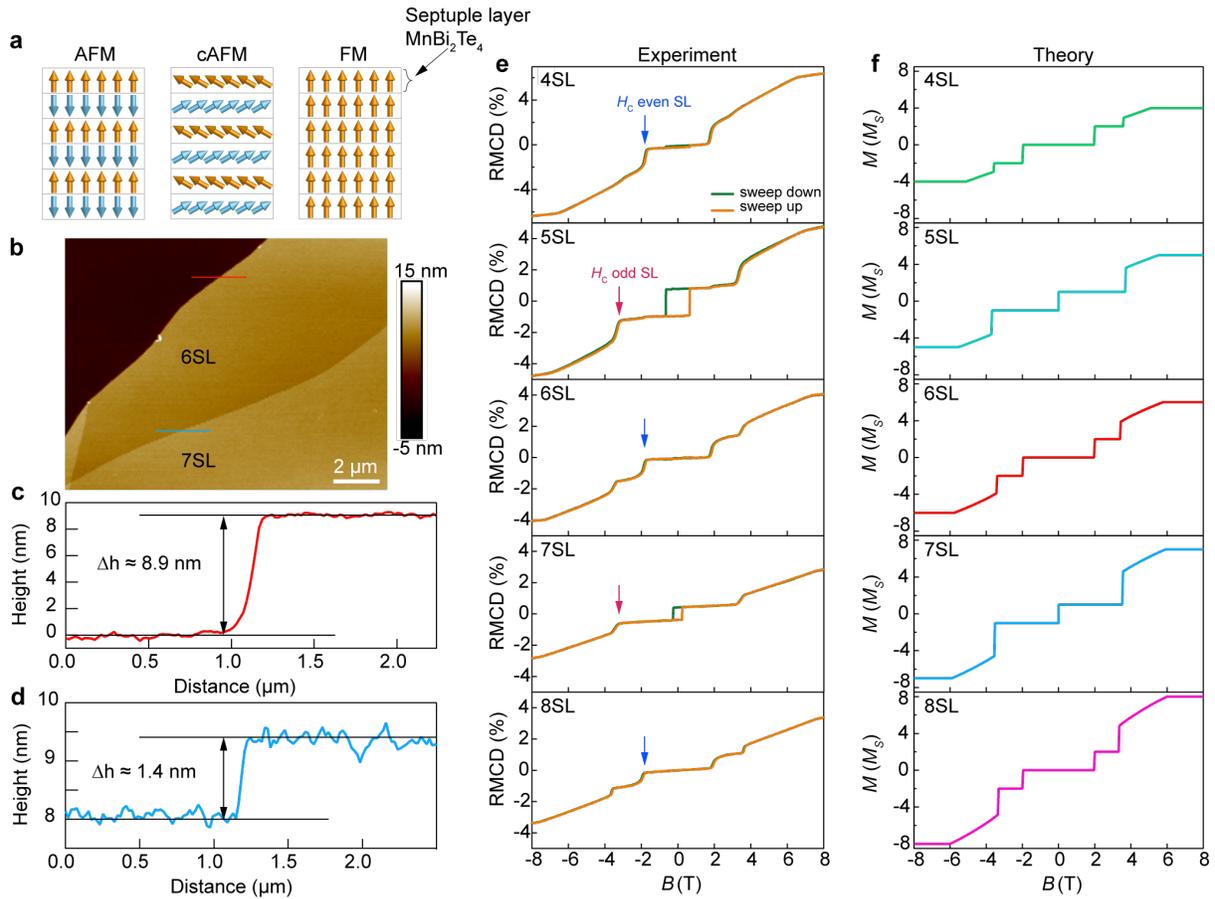

**Figure 1 | Layer dependent magnetic order in thin flakes of MnBi$_2$Te$_4$ a**, Schematics of magnetic order evolution as a function of magnetic field. From left to right: antiferromagnetic (AFM), canted antiferromagnetic (cAFM), and ferromagnetic (FM) states, which consequently form under increase of perpendicular magnetic field *B*. **b**, Atomic force microscopy (afm) micrograph of freshly exfoliated MnBi$_2$Te$_4$ flakes with thicknesses of 6 SL and 7 SL. **c & d**, height profiles across the lines, denoted with corresponding colors in (**b**). The data in (**d**) shows a single SL step size of 1.4 nm. **e**, Reflective magnetic circular dichroism (RCMD) measurements as a function of magnetic field for 4-SL to 8-SL flakes at *T* = 2 K. All samples are freshly exfoliated and not covered by PMMA. **f**, Theoretical modelling of the magnetization with respect to external magnetic field for 4-8 SL flakes.

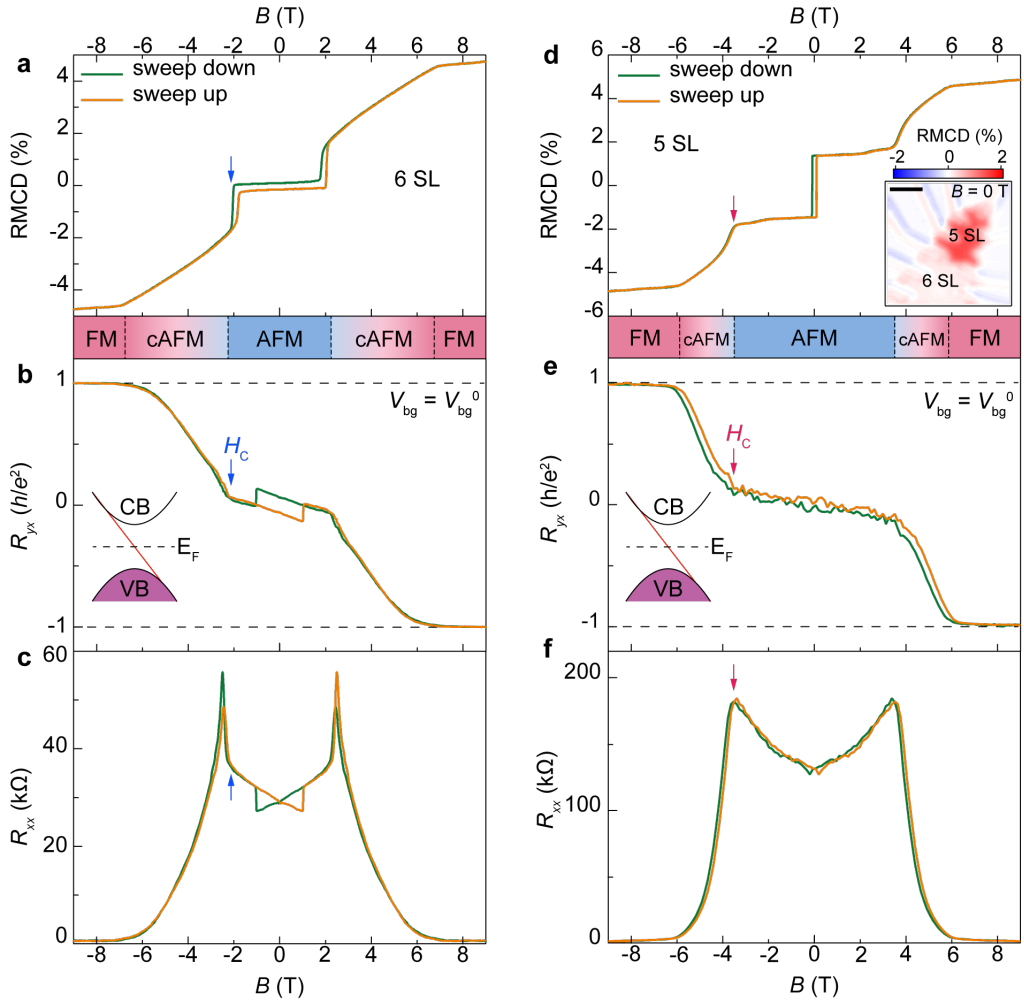

**Figure 2 | Correlation of magnetic and topological properties of even and odd SL MnBi$_2$Te$_4$ flakes.**
**a**, RMCD signal, **b**, anti-symmetrized Hall resistance, $R_{yx}$, **c**, symmetrized longitudinal resistance, $R_{xx}$, as a function of magnetic field in device 6SL-1 at $T = 2$ K. See Extended Data Fig. 1 for the raw data. Transport and RMCD measurements were taken at $V_{bg} = V_{bg}^0 = 37$ V and $V_{bg} = 0$ V, respectively. **d**, RMCD signal, **e**, anti-symmetrized Hall resistance, $R_{yx}$, **f**, symmetrized longitudinal resistance, $R_{xx}$, as a function of magnetic field. Transport measurements were taken at $V_{bg} = V_{bg}^0 = 44.4$ V. Inset in (**d**) is the RMCD map taken at zero magnetic field, showing a single magnetic domain for the 5SL-1 device. Scale bar in the inset: 10 μm. Insets in (**b**) & (**e**) depict the chiral edge state dispersion in the FM state.

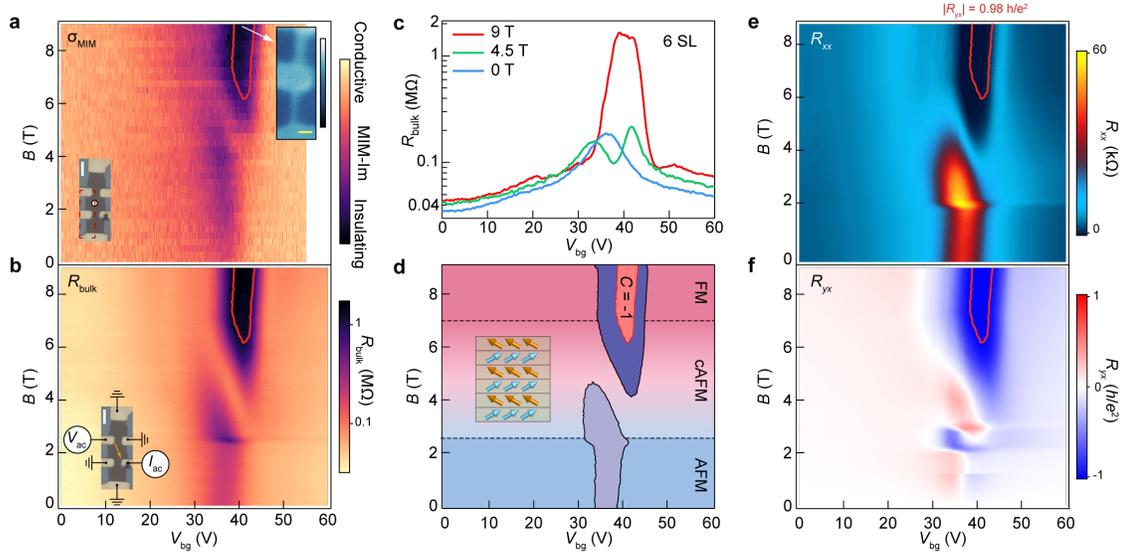

**Figure 3 | Probing band crossing during the topological phase transition and imaging of Chern gap in a 6-SL device. a**, The $V_{bg}$-$B$ map of the MIM-Im signal measured inside the bulk (indicated by the white dot in the lower inset, scale bar: 5 µm), labelled as $\sigma_{MIM}$, taken at $T = 3$ K. Upper inset (scale bar: 2 µm): spatial MIM scan of the device region outlined by the red dotted rectangle in the lower inset, taken in the Chern insulator state at $B = 9$ T and $T = 3$ K. **b,** The $V_{bg}$-$B$ map (log scale) of the bulk resistance, $R_{bulk}$, taken at $T = 2$ K. Inset shows the measurement configuration. Scale bar: 5 µm. **c,** Linecuts of $R_{bulk}$ vs $V_{bg}$ at $B = 0$ T, 4.5 T, and 9 T, extracted from (**b**). **d,** Phase diagram of the topological phase transition. The contours of $R_{bulk}$ = 150 kΩ (black solid lines) indicate the regions of the bulk insulating states in both the AFM and FM states. The contour of $|R_{yx}| = 0.98$ $h/e^2$ (red solid line), determined from (**f**), indicates the $C = -1$ Chern insulator region. Inset: schematics of the cAFM phase during the topological phase transition. **e & f,** The $V_{bg}$-$B$ maps of (**e**) the raw longitudinal resistance, $R_{xx}$, and (**f**) the anti-symmetrized Hall resistance $R_{yx}$, taken at $T = 2$ K. The contour of $|R_{yx}| = 0.98$ $h/e^2$ is overlaid in (**a,b,d,e,f**). All data shown here are from device 6SL-2.

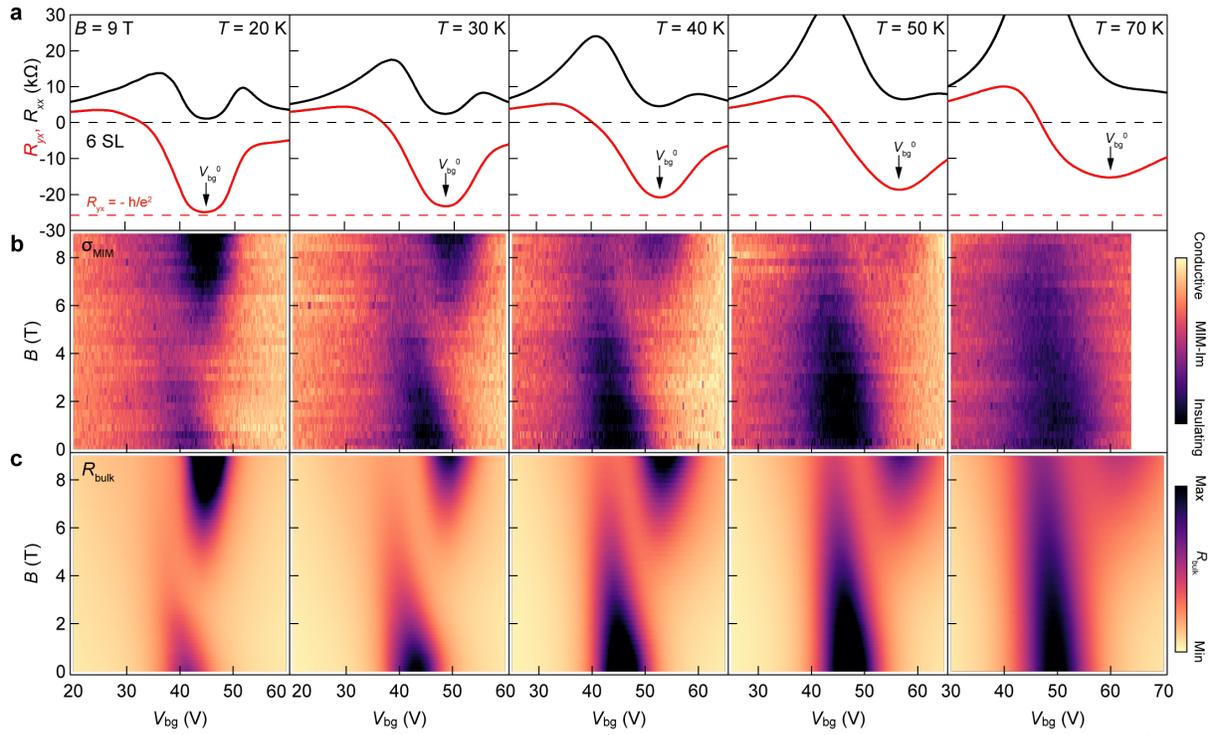

**Figure 4 | Temperature dependence of the band crossing and Chern insulator state.** All data shown here are from device 6SL-2. **a**, $R_{xx}$ (symmetrized) and $R_{yx}$ (anti-symmetrized) vs $V_{bg}$ curves at $B = 9$ T and selected temperatures. **b & c**, The corresponding $V_{bg}$-$B$ maps of (**b**) the bulk MIM-Im signal and (**c**) $R_{bulk}$. The colorscale for each panel is adjusted to highlight the band crossing feature.

# Extended Data Figures

# Intertwined Topological and Magnetic Orders in Atomically Thin Chern Insulator MnBi$_2$Te$_4$

**Authors**: Dmitry Ovchinnikov[1*], Xiong Huang[2,3*], Zhong Lin[1*], Zaiyao Fei[1*], Jiaqi Cai[1], Tiancheng Song[1], Minhao He[1], Qianni Jiang[1], Chong Wang[4], Hao Li[5], Yayu Wang[6], Yang Wu[7], Di Xiao[4], Jiun-Haw Chu[1], Jiaqiang Yan[8,9§], Cui-Zu Chang[10§], Yong-Tao Cui[2§], Xiaodong Xu[1,11§]

**Affiliations:**
[1]Department of Physics, University of Washington, Seattle, Washington 98195, USA.
[2]Department of Physics and Astronomy, University of California, Riverside, California 92521, USA
[3]Department of Materials Science and Engineering, University of California, Riverside, California 92521, USA
[4]Department of Physics, Carnegie Mellon University, Pittsburgh, Pennsylvania 15213, USA
[5]School of Materials Science and Engineering, Tsinghua University, Beijing, 100084, P. R. China
[6]Department of Physics, Tsinghua University, Beijing 100084, P. R. China
[7]Department of Mechanical Engineering, Tsinghua University, Beijing 100084,P. R. China
[8]Materials Science and Technology Division, Oak Ridge National Laboratory, Oak Ridge, Tennessee 37831, USA.
[9]Department of Materials Science and Engineering, University of Tennessee, Knoxville, Tennessee 37996, USA.
[10]Department of Physics, The Pennsylvania State University, University Park, Pennsylvania 16802, USA
[11]Department of Materials Science and Engineering, University of Washington, Seattle, Washington 98195, USA.
*These authors contributed equally to the work.
§Correspondence to yanj@ornl.gov; cxc955@psu.edu; yongtao.cui@ucr.edu; xuxd@uw.edu

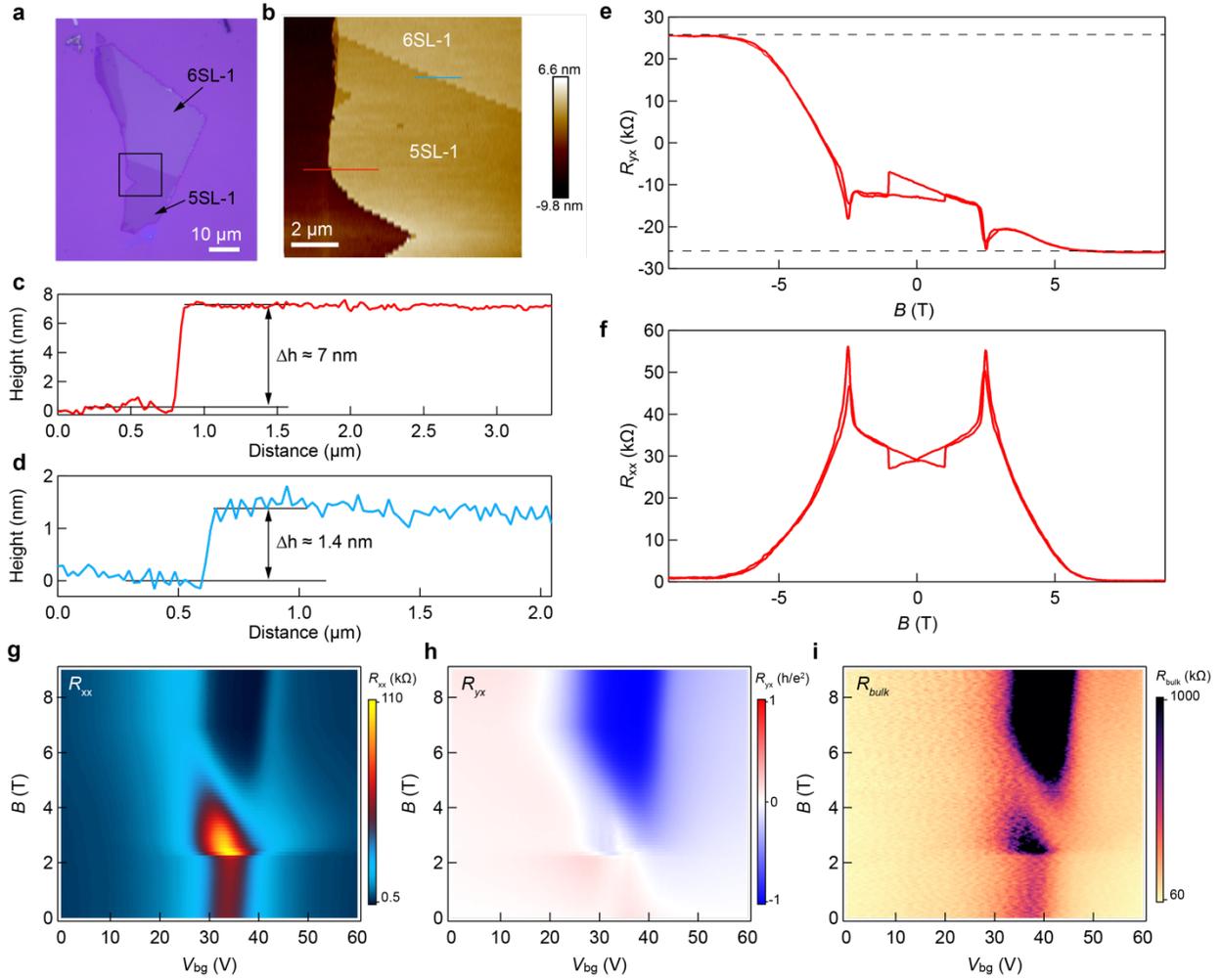

**Extended Data Figure 1 | Atomic force microscopy (afm) imaging and additional transport data for device 6SL-1 a,** Optical micrograph of MnBi$_2$Te$_4$ flake used for devices 5SL-1 and 6SL-1. **b,** afm image of the area indicated by the black rectangle in (**a**). **c-d,** Height profiles along the two lines outlined in (**b**), indicating a thickness of 7 nm for 5SL-1 and a monolayer step of 1.4 nm between 5SL-1 and 6SL-1. **e,** $R_{yx}$ raw data and **f,** $R_{xx}$ raw data, corresponding to Figs. 2b and 2c of the main text at $V_{bg}$ = 37 V and $T$ = 2 K. Dashed lines in (**e**) indicate $R_{yx} = \pm h/e^2$. **g-i,** Maps of symmetrized $R_{xx}$, antisymmetrized $R_{yx}$, $R_{bulk}$ as a function of $V_{bg}$ and $B$, respectively. Data is taken at $T$ = 2 K. $R_{bulk}$ is taken at different date and the gate range shift slightly compared to $R_{xx}$ and $R_{yx}$.
15

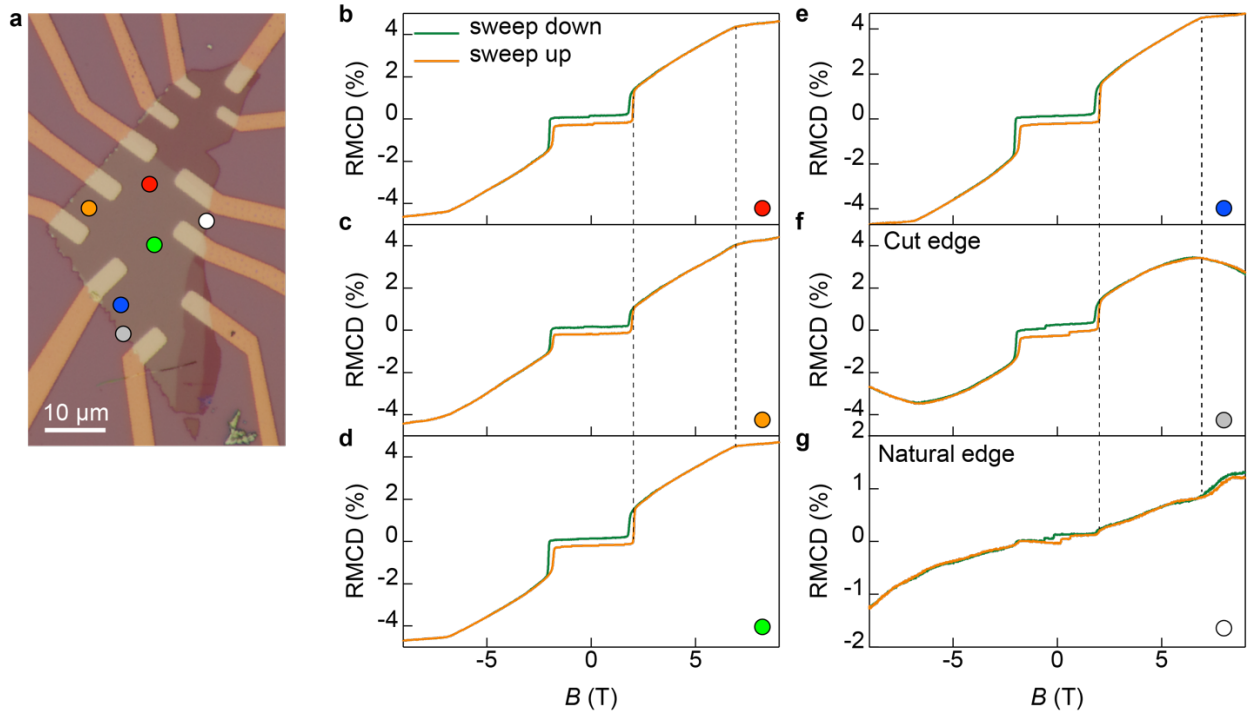

**Extended Data Figure 2 | RMCD for selected spots of 6SL-1. a,** Optical micrograph of devices 5SL-1 and 6SL-1. **b-g,** RMCD as a function of magnetic field $B$ for selected spots. We observed additional hysteresis transitions on both natural and pre-cut edges, as indicated by white and grey dot locations. Dotted lines indicate transitions between AFM, cAFM and FM states.

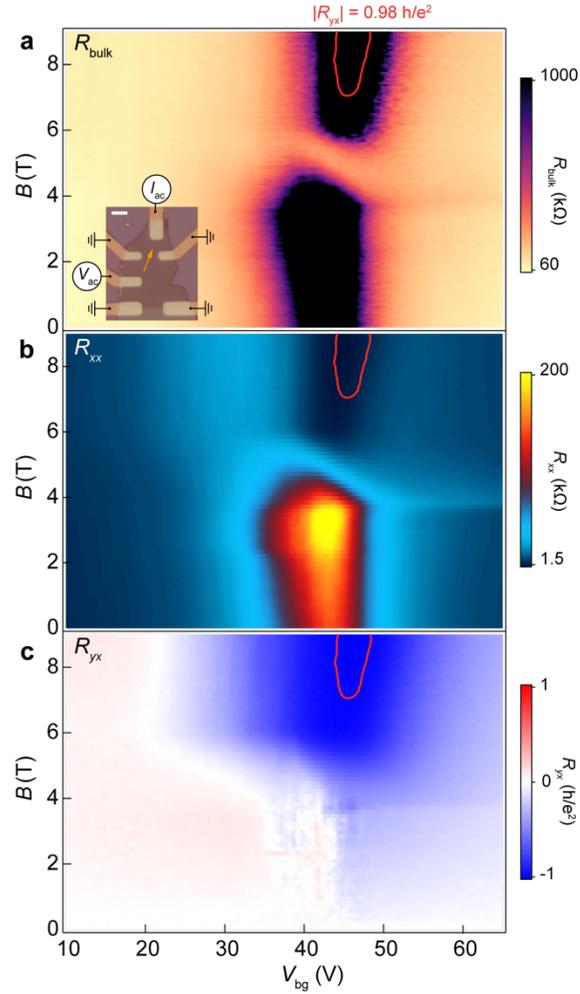

**Extended Data Figure 3 | Band crossing during the topological phase transition in a 5-SL device (5SL-1). a-c,** $V_{bg}$-$B$ maps of (**a**) the bulk resistance, $R_{bulk}$, (**b**) symmetrized longitudinal resistance, $R_{xx}$, and (**c**) anti-symmetrized Hall resistance, $R_{yx}$, respectively. The contour of $|R_{yx}| = 0.98\ h/e^2$ (red solid line) is overlaid in all three maps. Scale bar in the inset: 5 μm. All data shown here are from device 5SL-1 at $T = 2$ K.

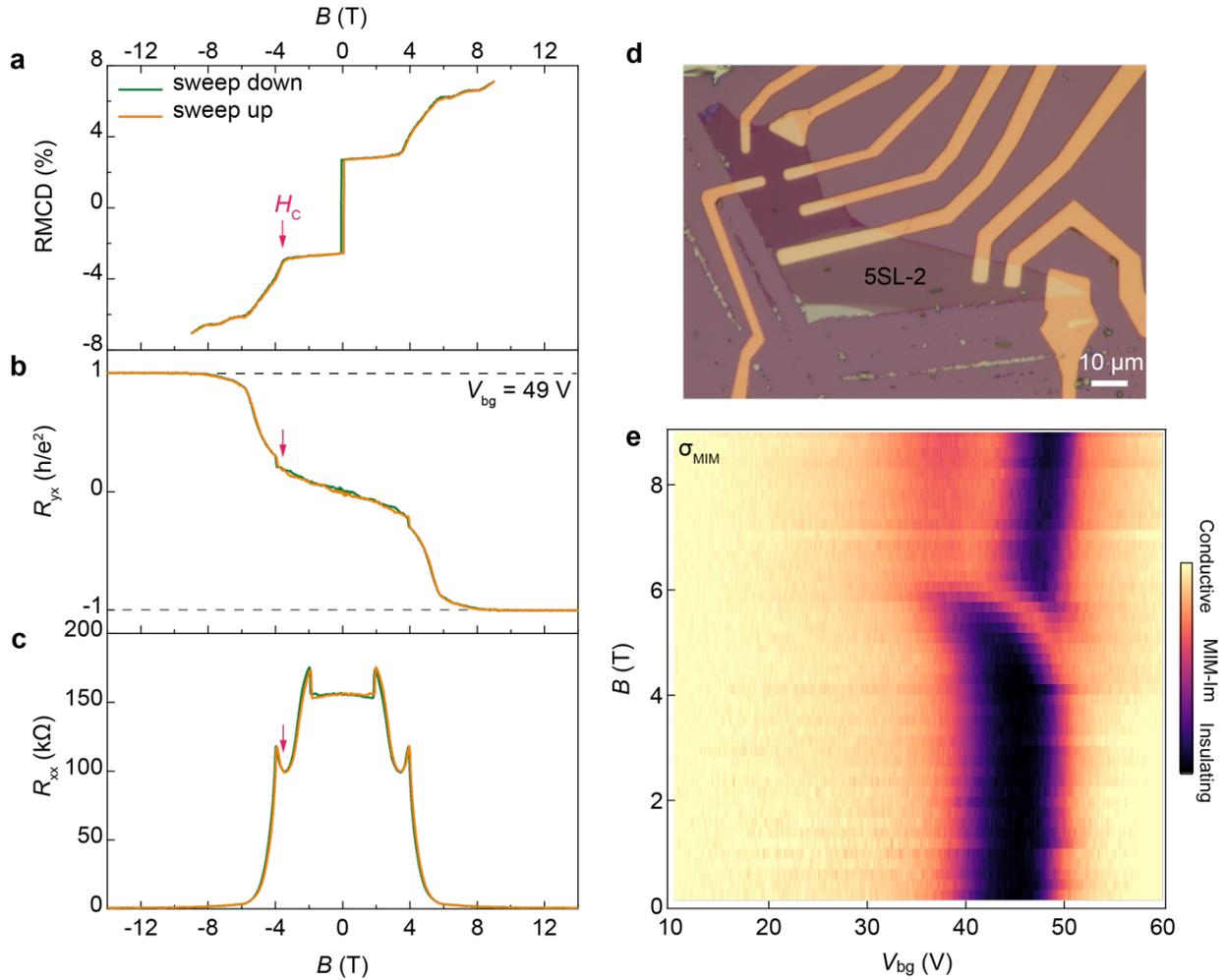

**Extended Data Figure 4 | Characterization of 5SL-2. a**, RMCD signal, **b**, anti-symmetrized Hall resistance, $R_{yx}$, **c**, symmetrized longitudinal resistance, $R_{xx}$, as a function of magnetic field in device 5SL-2 at $T = 2$ K. Transport and RMCD measurements were taken at $V_{bg} = V_{bg}^0 = 49$ V and $V_{bg} = 0$ V, respectively. Red arrows indicate critical field $H_C \sim 3.5$ T for spin-flop transition. **d,** Optical micrograph of device 5SL-2. **e,** MIM-Im signal as a function of $V_{bg}$ and $B$ at $T = 3$ K measured inside the bulk of 5SL-2.

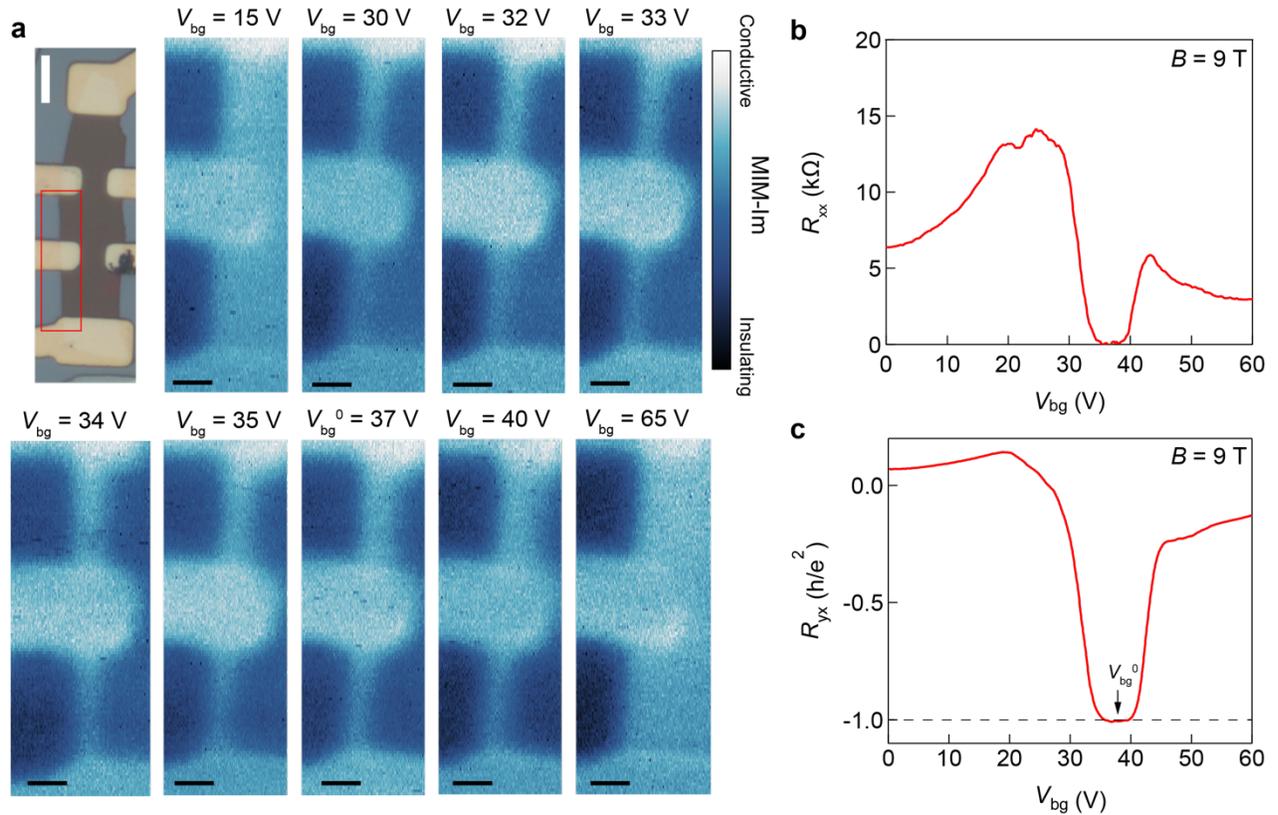

**Extended Data Figure 5 | Evolution of edge states in 6SL-2 as a function of gate voltage. a,** Optical micrograph of 6SL-2 (left top panel) and MIM images of edge states taken from the area outlined by red rectangle at various gate voltages. Scale bar is 5 μm on optical micrograph and 2 μm on MIM images. We define $V_{bg} = V_{bg}^0 = 37$ V as the center of the Chern insulator gap region. All MIM data is taken at $T = 3$ K and $B = 9$ T. **b, c,** $R_{xx}$ (symmetrized) and $R_{yx}$ (anti-symmetrized) of 6SL-2, as a function of $V_{bg}$ at $T = 2$ K and $B = 9$ T. We observe small spatial inhomogeneity comparing the gate dependence of the upper and lower regions. We attribute those doping inhomogeneities to the protective PMMA layer (see Methods) which is known to host charge disorders.

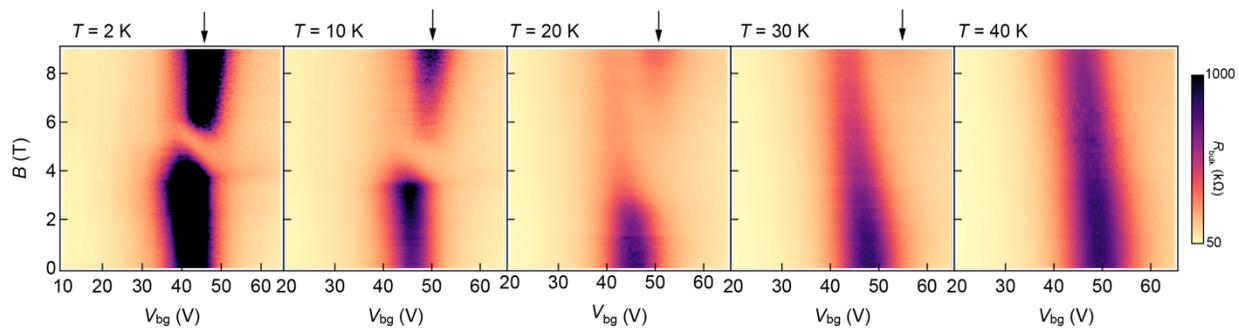

**Extended Data Figure 6 | Temperature dependence of $R_{bulk}$ for 5SL-1.** Chern insulator gap features can be resolved up to $T = 30$ K, indicated by black arrows.

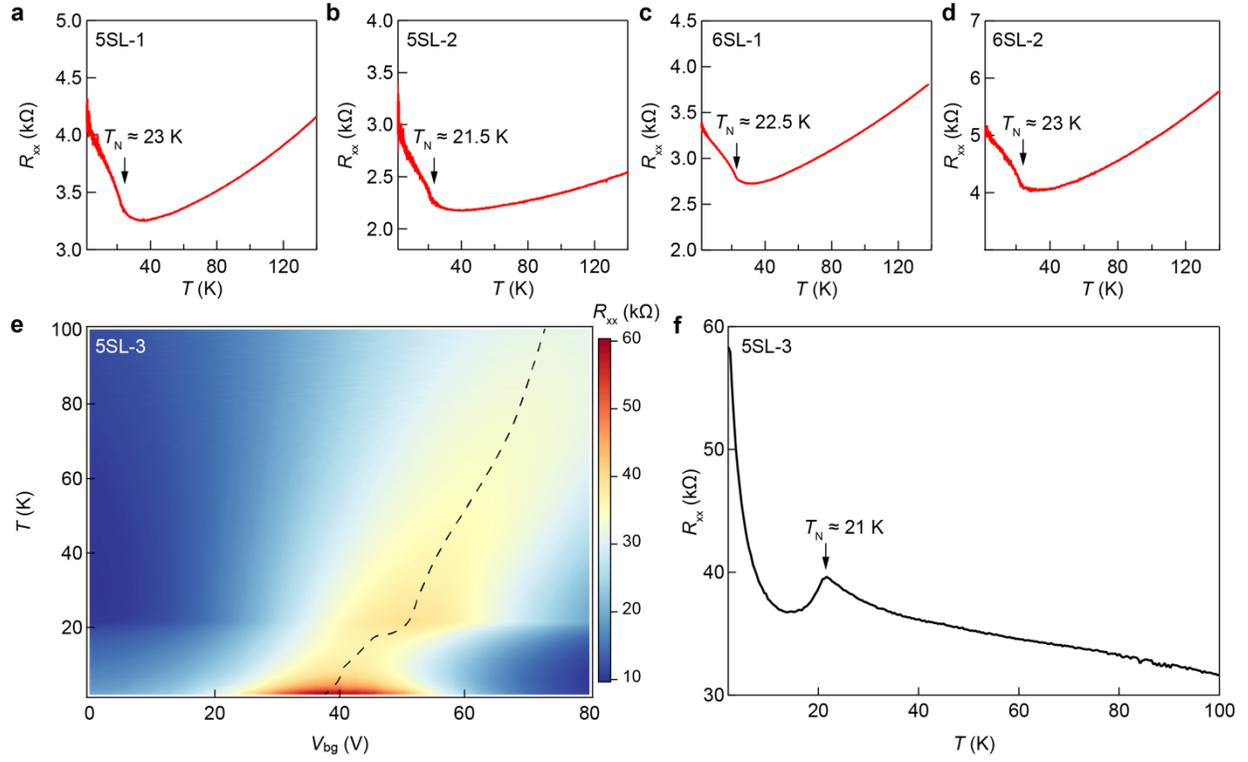

**Extended Data Figure 7 | Determination of Néel temperature $T_N$ from transport measurements. a-d,** Longitudinal resistance $R_{xx}$ as a function of temperature $T$ for 5SL-1, 5SL-2, 6SL-1, and 6SL-2. Kinks around 21-23 K indicate the transition from paramagnetic to antiferromagnetic state. **e,** Longitudinal resistance $R_{xx}$ as a function of temperature T and back gate voltage $V_{bg}$ for device 5SL-3. **f,** Longitudinal resistance $R_{xx}$ as a function of temperature tracked at the CNP as indicated by the dashed line in (e). All data is taken at magnetic field $B = 0$ T.

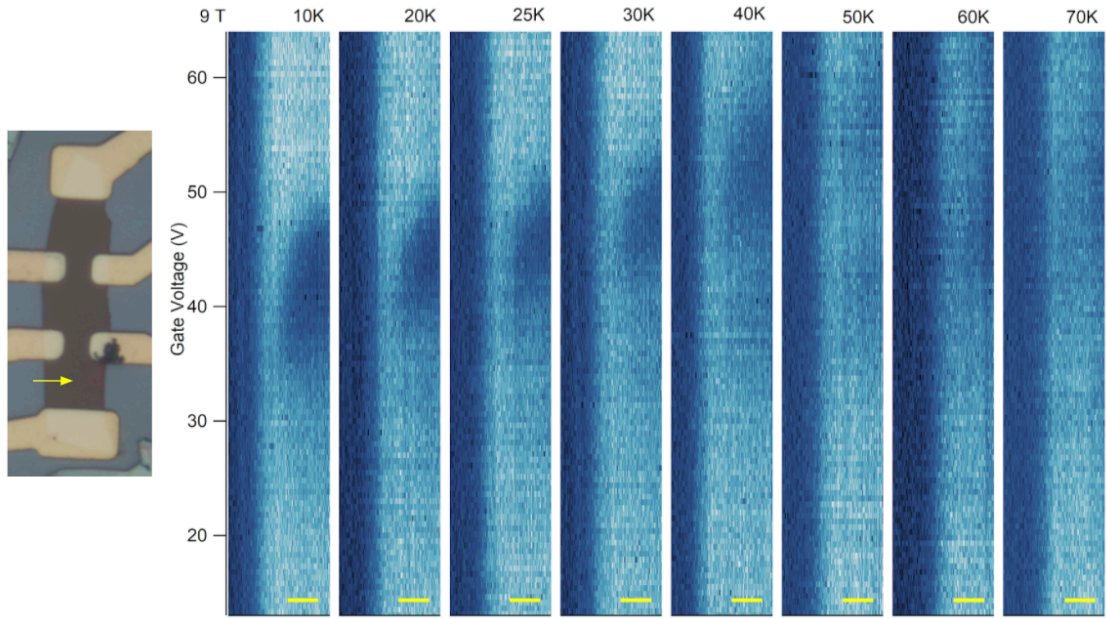

**Extended Data Figure 8 | Temperature dependence of MIM-Im line scans at *B* = 9 T in device 6SL-2.** The data is taken by repeatedly scanning along the same line across the sample edge indicated by the arrow in the optical image while tuning the gate voltage. All scale bars: 2 μm.